\documentstyle[aps,prb,multicol]{revtex}
\begin{document}
\title{\bf Josephson junction between anisotropic superconductors}
\author{Roman G. Mints}
\address{School of Physics and Astronomy, Raymond and Beverly Sackler
Faculty of Exact Sciences, \\Tel Aviv University, Tel Aviv 69978,
Israel}
\author{Vladimir G. Kogan}
\address{Ames Laboratory - DOE and Department of Physics, Iowa State
University, Ames Iowa 50011 }
\date{\today}
\maketitle
\begin{abstract}
The sin-Gordon equation for Josephson junctions with arbitrary
misaligned anisotropic banks is derived. As an application, the problem
of Josephson vortices at twin planes of a YBCO-like material is
considered. It is shown that for an arbitrary orientation of these
vortices relative to the crystal axes of the banks, the junctions
should experience a mechanical torque which is evaluated. This torque
and its angular dependence may, in principle, be measured in small
fields, since the flux penetration into twinned crystals begins with
nucleation of Josephson vortices at twin planes.
\end{abstract}
\begin{multicols}{2}
\narrowtext
\section{Introduction}
The high-$T_c$ superconductors are often used as polycrystals made of
anisotropic grains touching each other with various degrees of
crystallographic misalignment. As a rule, the contacts have Josephson
properties.  Hence, physical characteristics of Josephson junctions
with misaligned anisotropic banks are of considerable interest. To our
knowledge, the only publications on this subject were by one of us and
were concerned with a relatively simple situation of a junction between
perfectly aligned anisotropic superconductors.\cite{Mints1,Mints2}
\par
In this paper we consider two misaligned grains of the same uniaxial
material forming a Josephson contact. The material is characterized by
the dimensionless ``mass tensor" with eigenvalues $m_a=m_b<m_c$ which
are normalized so that $m_a^2 m_c=1$.  The mass ratios are defined as
$m_i/m_k=\lambda^2_i/\lambda_k^2$, so that the actual penetration
depths can be written as $\lambda_i=\lambda\sqrt{m_i}$ where
$\lambda=(\lambda^2_a\lambda_c)^{1/3}$.
\par
The paper is organized as follows. We first consider aligned banks not
only to establish notation and to demonstrate the approach; we have in
mind a material of YBCO family which usually contains sets of nearly
parallel twinning planes having Josephson properties. We show that the
gauge invariant phase difference still satisfies the sine-Gordon
equation, however, the squared Josephson length acquires tensor
properties which are discussed in detail. Next, the case of a general
misalignment is treated and the tensor for the squared Josephson length
is expressed in terms of mass tensors of the banks and their
misalignment.
\par
In twinned materials of the YBCO kind, in increasing from zero magnetic
field, the first vortices entering material are usually situated on the
twin planes and have Josephson properties. We consider the problem of a
Josephson vortex between anisotropic grains and evaluate the torque
which should act on the system of such vortices and via them on the
sample. Although this torque is small, given recent developments in
improving sensitivity of the torque magnetometry,\cite{Rossel} it is a
measurable quantity, from which the Josephson characteristics of grain
boundaries can, in principle, be extracted.
\par
\section{Main equations}
\subsection{Aligned banks}
Let us first consider the simplest possible situation: a Josephson
junction with superconducting banks made of the {\it same} anisotropic
material; $m_{ik}$ are the same on both sides and the banks are
perfectly aligned. Moreover, let the axis $y$ perpendicular to the
junction plane $xz$ be the principal direction, say $b$. Let us assume
further that a Josephson vortex is directed at an angle $\theta$
relative to the crystal axis $c$. We choose the $z$ axis along the
vortex.
\par
The coordinates $x,z$ are related to $a,c$:
$x=a\cos\theta +c\sin\theta$ and $z=-a\sin\theta + c\cos\theta $. Then
the nonzero components of the mass tensor in the frame $xyz$ are:
\begin{eqnarray}
m_{xx}&=&m_a\cos^2\theta +m_c\sin^2\theta\,,
\nonumber\\
m_{zz}&=&m_a\sin^2\theta +m_c\cos^2\theta\,,
\label{m_ik}\\
m_{xz}&=&(m_c-m_a)\cos\theta\sin\theta\,,\quad m_{yy}=m_a\,.
\nonumber
\end{eqnarray}
Note a useful property
\begin{equation}
m_{xx}m_{zz}-m_{xz}^2=m_am_c
\label{property}
\end{equation}
which follows from det$(m_{ik})=m_a^2m_c$.
\par
We proceed by calculating the London energy $F_L$ associated with a
Josephson vortex along $z$ in terms of the gauge invariant phase
$\varphi$. We then minimize the total energy $F_L+F_J$ where
\begin{equation}
F_J=\frac{\phi_0j_c}{2\pi c}\int_{-\infty}^{\infty}dx\, (1-\cos
\varphi)\,,
\label{E_j}
\end{equation}
with respect to $\varphi$ to obtain the tunneling current $j_y$ in
terms of $\varphi$ along with the equation governing the phase; here
$\phi_0$ is the flux quantum and $j_c$ is the Josephson critical current
density.
\par
For a straight vortex along $z$, all gauge invariant quantities are $z$
independent: the field components $h_{x,z}$ are functions of $x$ and
$y$, $h_y=0$, while $\varphi$ depends only on $x$.
\par
We begin with the general expression for the London energy \cite{K81}
\begin{equation}
F_L={1\over 8\pi}
\int (h^2+\frac{4\pi\lambda^2}{c}m_{ik}j_i{\rm curl}_k{\bf h})dV
\label{L}
\end{equation}
and integrate the kinetic part by parts to obtain
\begin{equation}
F_L=\frac{ \lambda^2}{2c}m_{ik}e_{klm}\oint j_i h_m dS_l\,,
\label{L1}
\end{equation}
where $d{\bf S}=\pm dx\,{\hat {\bf y}}$ is an area element on two
junction sides. This gives
\begin{equation}
F_L=\frac{\lambda^2}{2c}\int_{-\infty}^{\infty}dx\,\{H_x\,[m_{zk}j_k]-
H_z\, [m_{xk}j_k]\}
\label{L2}
\end{equation}
per unit length in the $z$ direction. Hereafter $H_{x,z}(x)$ are the
field components at the junction plane $y=0$; $y$ is directed from the
side 1 to the side 2 of the junction. The symbol $[...]$ denotes the
difference of a quantity on two sides, e.g.,
$[m_{xy}]=m_{xy}^{(2)}-m_{xy}^{(1)}$.
\par
We now utilize the London equations
\begin{equation}
A_i+{\phi_0\over 2\pi}{\partial\chi\over\partial x_i}
= -{4\pi\lambda^2\over c}m_{ik} j_k\,,
\label{London}
\end{equation}
where $\chi$ is the phase of the order parameter. Also, we use the
gauge $A_y=0$ and the continuity of the tangential component of the
vector potential to obtain
\begin{eqnarray}
&&m_{xx} [j_x]+m_{xz}[j_z]=-{c\phi_0\over 8\pi^2\lambda^2}
{d\varphi\over d\,x}\,,
\label{bc1}\\
&&m_{zx} [j_x]+m_{zz}[j_z]=0\,;
\label{bc2}
\end{eqnarray}
here we introduced the phase difference $\varphi =[\chi]$, utilized the
alignment, and the condition $[j_y]=0$. Note that for the uniform along
$z$ solutions, the discontinuities of the tangential currents $j_x$ and
$j_z$ at the junction are proportional to each other. Solving the
system (\ref{bc1}), (\ref{bc2}) we obtain:
\begin{equation}
[j_x]=-\frac{m_{zz} }{m_{xz} }[j_z]=
-{c\phi_0m_{zz}\over 8\pi^2\lambda^2m_am_c}\varphi\,^{\prime}(x)\,
\label{*}
\end{equation}
where the property  (\ref{property}) has been used.
\par
With the help of Eqs. (\ref{bc1}) and (\ref{bc2}) the London energy
takes the form:
\begin{equation}
F_L=\frac{\phi_0}{16\pi^2}\int_{-\infty}^{\infty}dx\,
H_z(x)\, \varphi^{\,\prime}(x)
\label{L-energy}
\end{equation}
To make further progress one has to turn to the field and current
distributions in the banks. Assume that
\begin{equation}
\lambda \ll \lambda_J\,,
\end{equation}
where the length $\lambda_J$ is the characteristic distance at which
$\vec{H}$ changes in the junction plane. In this case one can consider
the field penetration into the banks as the standard Meissner problem
of a uniform ``applied'' field $\vec{H}=\{H_a,H_c\}$ penetrating in two
half-spaces $y>0$ and $y<0$. Since $h_a=H_a\exp(-|y|/\lambda_c)$ and
$h_c=H_c\exp(-|y|/\lambda_a)$, we obtain
\begin{eqnarray}
h_x&=&H_ae^{-|y|/\lambda_c}\cos\theta
+ H_ce^{-|y|/\lambda_a}\sin\theta\,,
\nonumber\\
h_z&=&-H_ae^{-|y|/\lambda_c}\sin\theta
+ H_ce^{-|y|/\lambda_a}\cos\theta\,,
\label{**_}
\end{eqnarray}
where $\lambda_{a,c}=\lambda\sqrt{m_{a,c}}$. In terms of $H_{x,z}$ we
have
\begin{eqnarray}
h_x&=&H_x\Big(e^{-|y|/\lambda_c}\cos^2\theta+
e^{-|y|/\lambda_a}\sin^2\theta\Big)
\nonumber\\ &+&H_z\Big (e^{-|y|/\lambda_a}
-e^{-|y|/\lambda_c}\Big )\sin\theta\cos\theta\,,
\label{hx} \\
h_z&=&H_x\Big (e^{-|y|/\lambda_a} -e^{-|y|/\lambda_c}\Big )
\sin\theta\cos\theta
\nonumber\\
&+&H_z\Big (e^{-|y|/\lambda_c}\sin^2\theta+
e^{-|y|/\lambda_a}\cos^2\theta\Big )\,.
\label{hz}
\end{eqnarray}
The field is continuous at the junction plane $y=0$ whereas the
tangential currents $j_{x,z}$ are not. One finds:
\begin{eqnarray}
{4\pi\over c}[j_x]&=&H_x\Big ({1\over\lambda_c}
-{1\over\lambda_a} \Big )\sin 2\theta
\nonumber\\
&-&2H_z\Big ({1\over\lambda_c} \sin^2\theta +
{1\over\lambda_a}\cos^2\theta\Big )
\label{jx}\,,\\
{4\pi\over c}[j_z] &=&2H_x\Big ({1\over\lambda_c} \cos^2\theta
+ {1\over\lambda_a} \sin^2\theta\Big )
\nonumber\\
&+& H_z\Big ({1\over\lambda_c}-{1\over\lambda_a}\Big )\sin 2\theta\,.
\label{jz}
\end{eqnarray}
Note that the coefficients of $H_{x,z}$ in these expressions do not
reduce to components $m_{ik}$ of Eq. (\ref{m_ik}) because they contain
$1/\sqrt{m_{a,c}}$ instead of $m_{a,c}$. It is convenient to introduce
another tensor $\mu_{\alpha\beta}$ ($\alpha,\beta=x,z$) with eigenvalues
$1/\sqrt{m_{a,c}}\,$ (to obtain $\mu_{\alpha\beta}$ replace $m_{a,c}$
in Eq. (\ref{m_ik}) with $1/\sqrt{m_{a,c}}$). Then, Eqs. (\ref{jx}),
(\ref{jz}) take a compact form:
\begin{eqnarray}
{2\pi\over c}[j_x]\lambda &=& H_x \mu_{xz} -H_z \mu_{xx}
\label{jx1}\,,\\
{2\pi\over c}[j_z]\lambda &=&H_x \mu_{zz} - H_z \mu_{xz}\,.
\label{jz1}
\end{eqnarray}
Since the ratio $[j_x]/[j_z]$ is fixed by Eq. (\ref{bc2}), we obtain a
fixed ratio $H_x/H_z$:
\begin{equation}
H_x(\mu_{xz}m_{xz}+\mu_{zz}m_{zz})=H_z(\mu_{xx}m_{xz}+\mu_{xz}m_{zz})
\label{Hx/Hz}
\end{equation}
Clearly, the transverse field $H_x$ vanishes if the vortex axis
coincides with either $a$ or $c$.
\par
We now go back to Eq. (\ref{*}) and obtain using (\ref{jx1}) and
(\ref{Hx/Hz}):
\begin{equation}
\frac{\phi_0}{4\pi\tilde{\lambda}}\varphi\,^{\prime}(x)=H_z\,,
\label{21}
\end{equation}
where
\begin{equation}
\tilde{\lambda}=\frac{\lambda }{
\mu_c\sin^2\theta+\mu_a\cos^2\theta}={\lambda\over \mu_{xx}}\,.
\label{l-tild}
\end{equation}
\par
Thus, the London energy
\begin{equation}
F_L=\frac{\phi_0^2}{64\pi^3\tilde{\lambda}}\int_{-\infty}^{\infty}dx\,
(\varphi^{\,\prime})^2\,.
\label{L-en}
\end{equation}
We complement this with the Josephson energy (\ref{E_j}) and minimize
the sum with respect to the phase to obtain the result of
Ref. \onlinecite{Mints1}:
\begin{equation}
\lambda_J^2\mu_{xx}\varphi^{\,\prime\prime}= \sin\varphi\,,
\label{sineG}
\end{equation}
where the length $\lambda_J$ is defined as
\begin{equation}
\lambda_J =\sqrt{c\phi_0\over 16\pi^2\lambda j_c}.
\label{jl}
\end{equation}
It follows from Eq. (\ref{sineG}) that the Josephson length is
\begin{equation}
\Lambda_J(\theta)=\lambda_J\sqrt{\mu_{xx}(\theta)}\,,
\label{j1a}
\end{equation}
where
\begin{equation}
\mu_{xx}=k^{-2/3}\,\sin^2\theta + k^{1/3}\,\cos^2\theta,
\label{jl1}
\end{equation}
and $k=\lambda_c/\lambda_a$ is the anisotropy parameter. Note that
$\lambda_J$ has now meaning of an {\it average} Josephson length. We see
therefore that even in the case of aligned banks considered here, the
Josephson length $\Lambda_J(\theta)$ depends on the orientation of the
Josephson vortex within the junction plane $ac$. In particular, we have
\begin{equation}
{\Lambda_J(0)\over\Lambda_J(\pi/2)}=\sqrt{k}.
\end{equation}
\par
Using Eqs. (\ref {E_j}) and (\ref{L-en}) and the classical vortex solution of
Eq. (\ref{sineG}),  $\varphi (x)=4\tan^{-1}[\exp (x/\Lambda_J)]$,
we calculate the line energy of the Josephson vortex
\begin{equation}
\epsilon = {1\over 4\pi^3 k^{1/3}}\,{\phi_0^2\over\lambda\lambda_J}
\sqrt{\sin^2\theta + k\cos^2\theta}.
\label{ener}
\end{equation}
\par
\subsection{Misaligned banks}
Consider now a simple misalignment: the banks are made of
identical materials with the axis $b$ on both sides along $y$ as before, but
the $a,c$ axes are rotated round $b$ through different angles on two junction
sides. We again look for vortex solutions uniform along the vortex
direction $z$. The mass tensors on two sides are given by Eq.
(\ref{m_ik}), however, the angles $\theta_1\ne\theta_2$. Equation
(\ref{L2}) still holds, but in Eqs.(\ref{bc1}) and (\ref{bc2}) the
components of $m_{ik}$ cannot be taken out of the square brackets
denoting the differences between quantities enclosed on the two sides:
\begin{eqnarray}
&&[m_{xi}j_i] =-{c\phi_0\over 8\pi^2\lambda^2}{d\varphi \over d\,x}\,,
\label{bc'}\\
&&[m_{zi} j_i] =0\,;
\label{bc''}
\end{eqnarray}
note that the summation index $i=x,y,z$ can be replaced with the two
dimensional (2D) $\alpha =x,z$ because $m_{\alpha y}=0$ on both sides.
We now see that Eq. (\ref{L-energy}) holds, too.
\par
The field distribution in the bank $y>0$ is obtained replacing $|y|$
and $\theta$ in Eqs. (\ref{hx}), (\ref{hz}) with $y$ and $\theta_2$,
whereas for $y<0$, $|y|\rightarrow -y$ and $\theta \rightarrow
\theta_1$. Then, one obtains for the currents at the junction side
$y=+0$:
\begin{eqnarray}
4\pi\lambda j_x^+  &=& c(H_x \mu_{xz}^+ -H_z \mu_{xx}^+)
\label{jx(0)}\,,\\
4\pi\lambda j_z^+ &=&c(H_x \mu_{zz}^+ - H_z \mu_{xz}^+)\,.
\label{jz(0)}
\end{eqnarray}
These relations can be written in a compact form with the help of the
two-dimensional unit antisymmetric tensor $e_{\alpha\beta}$:
\begin{equation}
4\pi\lambda j_{\alpha}^+/c = -
\mu_{\alpha\beta}^+e_{\beta\gamma}H_{\gamma}\,.
\label{j+}
\end{equation}
For the other side of the junction at $y=-0$ we have
\begin{equation}
4\pi\lambda j_{\alpha}^-/c =
\mu_{\alpha\beta}^-e_{\beta\gamma}H_{\gamma}\,.
\label{j-}
\end{equation}
\par
To proceed, we rewrite the system (\ref{bc'}), (\ref{bc''}) as
\begin{equation}
[m_{\alpha\beta}j_{\beta}] =-{c\phi_0\over 8\pi^2\lambda^2}
\varphi^{\,\prime}(x)\delta_{x\alpha}\,,
\label{mbc}
\end{equation}
substitute here Eq. (\ref{j+}), (\ref{j-}), and obtain
\begin{equation}
p_{\alpha\beta}H_{\beta} ={\phi_0\over 2\pi\lambda}
\varphi^{\,\prime} (x) \delta_{x\alpha}\,.
\label{system}
\end{equation}
Here
\begin{equation}
p_{\alpha\beta}=\{m_{\alpha\gamma}\mu_{\gamma\delta}\}e_{\delta\beta}\,
\label{p}
\end{equation}
and $\{...\}$ denotes the sum of a quantity enclosed taken on two
junction sides.
\par
The system of two equations (\ref{system}) can be solved for $H_{x,z}$:
\begin{equation}
H_z=-{\phi_0\over 2\pi\lambda}\,\frac{p_{zx}}{{\rm det}(p_{\alpha\beta})}
\varphi^{\,\prime}(x)\,.
\label{Hz}
\end{equation}
This coincides with Eq. (\ref{21}) in which
\begin{equation}
\tilde{\lambda}=-\lambda\,\frac{{\rm det}(p_{\alpha\beta})}{2p_{zx}}\,.
\label{l-tild1}
\end{equation}
To have the following equations in a more symmetric form, we introduce
a tensor
\begin{equation}
q_{\mu\nu}=-e_{\mu\alpha}p_{\alpha\nu}=
-e_{\mu\alpha}\{m_{\alpha\beta}\mu_{\beta\gamma}\} e_{\gamma\nu}
\label{q}
\end{equation}
instead of the pseudotensor ${\hat p}$; a compelling reason for this is
given below. Then
\begin{equation}\tilde{\lambda}=\lambda\,
\frac{{\rm det}(q_{\alpha\beta})}{2q_{xx}} \,
\label{l-tild2}
\end{equation}
and proceeding as above we obtain the sine-Gordon equation for the
phase:
\begin{equation}
\Lambda_{xx}^2
\frac{d^2\varphi}{dx^2}=\sin\varphi\,,\qquad
\Lambda_{xx}^2=\lambda_J^2\frac{2q_{xx}}{{\rm det}(q_{\alpha\beta})}
\label{Lambda}
\end{equation}
\par
For aligned banks ($\theta_1=\theta_2$) this coincides with Eq.
(\ref{l-tild}). To see this, note that in this case ${\hat m}$ and
${\hat \mu}$ are the same on both sides and
$\{m_{\alpha\beta}\mu_{\beta\gamma}\}=2m_{\alpha\beta}\mu_{\beta\gamma}$.
Then, $p_{zx}=-2m_{z\gamma}\mu_{\gamma z}=-2(m_a\mu_a\sin^2\theta
+m_c\mu_c\cos^2\theta)$ and ${\rm det}(p_{\alpha\beta})= 4m_a\mu_am_c\mu_c$
(choose the principal axes to calculate this invariant).
\par
\subsection{General equation for the phase}
We derive now an equation for the phase difference $\varphi (x,z)$ which
holds at the junction plane $xz$ with no reference to a particular
vortex solution. Since we no longer have uniformity in the $z$
direction, Eq. (\ref{L2}) takes the form
\begin{equation}
F_L=\frac{\lambda^2}{2c}\int dx\,dz\,(H_x\,[m_{zk}j_k]-
H_z\, [m_{xk}j_k])\,.
\label{L3}
\end{equation}
Using the London relations (\ref{London}) we obtain instead of
(\ref{mbc}):
\begin{equation}  [m_{\alpha\beta}j_{\beta}] =-{c\phi_0\over
8\pi^2\lambda^2}\frac{\partial\varphi }{\partial x_{\alpha}}\,.
\label{40}
\end{equation}
The current distributions are still given by Eqs. (\ref{j+}), (\ref{j-});
the system (\ref{system}) now reads:
\begin{eqnarray}
p_{xx} H_x +p_{xz}H_z&=&  {\phi_0\over
2\pi\lambda}\frac{\partial\varphi }{\partial x }\,,\\
p_{zx} H_x +p_{zz}H_z &=&  {\phi_0\over
2\pi\lambda}\frac{\partial\varphi }{\partial z}
\label{system1}
\end{eqnarray}
with the tensor $p_{\alpha\beta}$ defined in Eq. (\ref{p}). Solve this
for $H_{x,z}$ and substitute to the energy (\ref{L3}):
\FL
\begin{eqnarray}
F_L=&-&{\phi_0^2\over
32\pi^3\lambda\,{\rm det}(p_{\alpha\beta})}\,\int dx\,dz\,
\Big [(p_{zz}-p_{xx})\frac{\partial\varphi }{\partial x}
\frac{\partial\varphi }{\partial z}+\nonumber\\
&+&p_{zx}\Big (\frac{\partial\varphi
}{\partial x}\Big )^2 - p_{xz}\Big (\frac{\partial\varphi
}{\partial z}\Big )^2  \Big ].
\label{EL}
\end{eqnarray}
Since the scalar of energy cannot depend on the choice of coordinates,
the coefficients by the derivatives of $\varphi$ must form a tensor. This
is the tensor $q_{\alpha\beta}$ introduced above in Eq. (\ref{q}):
$p_{xx}=q_{zx},\,\,\,
p_{xz}=q_{zz},\,\,\,p_{zx}=-q_{xx}$, and $p_{zz}=-q_{xz}$. We then have
\begin{equation}
F_L={\phi_0^2\over
32\pi^3\lambda\,{\rm det}(q_{\alpha\beta})}
 \int dx\,dz\,  q_{\alpha\beta} \frac{\partial\varphi }{\partial
x_{\alpha}}
\frac{\partial\varphi }{\partial x_{\beta}}\,.
\label{EL1}
\end{equation}
\par
Minimizing with respect to $\varphi$ the sum of this energy with
\begin{equation}
F_J=\frac{\phi_0j_c}{2\pi c}\int dx\,dz\,(1-\cos \varphi )\,,
 \label{F_j}
\end{equation}
one obtains
\begin{equation}
\Lambda_{\alpha\beta}^2
\frac{\partial^2\varphi}{\partial x_{\alpha}\partial
x_{\beta}}=\sin\varphi\,,\qquad
\Lambda_{\alpha\beta}^2=\lambda_J^2\frac{2q_{\alpha\beta}}{{\rm
det}(q_{\alpha\beta})}\,.
\label{result}
\end{equation}
\par
\section{Josephson vortices in flat samples}
\subsection{Infinite slab}
Consider an infinite thick slab with $z$ axis perpendicular to the slab
surface and with a set of parallel Josephson contacts at planes $y=0,\pm
d, \pm 2d,...$ Clearly, the system for which this model is relevant is a
platelet of YBCO with a set of parallel twins. Let the magnetic field
${\vec{\cal H}}$ be applied in the plane $xz$ of the junctions at an angle
$\alpha$ to the $z$ axis. The fields of our interest are small enough so
that only Josephson vortices are nucleated at the junctions while
Abrikosov vortices are absent. For a general tilt $\alpha$ of the
applied field, the vortices are tilted as well at an angle $\beta$ still
to be determined. We denote as $N$ the line density of Josephson
vortices at each junction so that the average magnetic induction is
\begin{equation}
 B={N\over d}\,\phi_0\,. \label{induction}
\end{equation}
\par
The macroscopic boundary conditions for the slab geometry are
\begin{equation}
 B_z={\cal H}_z\,,\qquad H_x={\cal H}_x\,. \label{bc}
\end{equation}
We utilize the first of these to obtain:
\begin{equation}
{\cal H}_z=B\cos\beta = \frac{N}{d}\,\phi_0\cos\beta\,,
\label{Bz}
\end{equation}
 whereas
\begin{equation}
B_x =B\sin \beta  = {\cal H}_z\tan\beta\,. \label{Bx}
\end{equation}
\par
Since we are interested in the limit of vanishing density $N$ of vortices, we
can disregard their interaction and write the
Helmholtz free energy as
\begin{equation}
 F= {N\over d}\,\epsilon ( \beta)\,, \label{F}
\end{equation}
where the line energy $\epsilon$ of a Josephson vortex is given in
Eq. (\ref{ener}). With the help of Eq. (\ref{Bz}) we write:
\begin{equation}
 F= {1\over 4\pi}\,{\cal H}_zH_0 \sqrt{k +\tan^2 \beta}\,. \label{F1}
\end{equation}
where
 \begin{equation}
  H_0={4\over \pi^2k^{1/3}}\,{\phi_0\over\lambda\lambda_J}.
\label{H0}
\end{equation}
\par
We now turn to the thermodynamic potential which is minimum in
equilibrium at a given $\vec{\cal H}$. Given the boundary conditions
(\ref{bc}), this potential is
 \begin{equation}
  G =F - \frac{1}{4\pi}\, B_xH_x\,. \label{Gibbs}
\end{equation}
Since $\delta F=\vec{H}\cdot \delta \vec{B}/4\pi$, we have
 \begin{equation}
\delta G = \frac{1}{4\pi}(H_z\delta B_z-B_x\delta H_x)\,,\label{dGibbs}
\end{equation}
i.e., the potential $G$ is minimum indeed at fixed $B_z,H_x$ or
${\cal H}_z,{\cal H}_x$. It is worth noting that in fact $G$ is the
Gibbs energy with a subtracted term $B_zH_z/4\pi={\cal H}_z^2/4\pi$
which is constant for the slab geometry.
\par
Using Eqs. (\ref{Gibbs}), (\ref{F1}), and (\ref{bc}) we  obtain
\begin{equation}
 G(\vec{\cal H}, \beta)= {{\cal H}_z\over 4\pi}\,(H_0 \sqrt{k +
\tan^2 \beta}-{\cal H}_x\tan \beta)\,.
\label{GG}
\end{equation}
Minimization with respect to $\tan \beta$ yields the direction of
 vortices:
\begin{equation}
 \tan^2 \beta= {k\,{\cal H}_x^2\over   H_0^2-{\cal H}_x^2} \,.
\label{tg}
\end{equation}
We observe that when the direction of the applied field changes from
$\alpha=0$ to $\alpha=\pi/2$, the vortices ($\vec{B}$) rotate from
$\beta=0$ to $\beta_{\rm max}$ such that
\begin{equation}
 \tan^2\beta_{\rm max}= {k\,{\cal H}_x^2\over  H_0^2-{\cal H}^2} \,.
\label{b-max}
\end{equation}
\par
It is of interest to see how the equilibrium vortex density changes
during rotation of the constant in value external field $\vec{\cal H}$:
\begin{equation}
B^2={\cal H}_z^2\tan^2\alpha +{\cal H}_z^2=
{\cal H}_z^2\frac{H_0^2+{\cal H}_x^2(k-1)}{H_0^2-{\cal H}_z^2}   \,.
\label{B^2}
\end{equation}
We see that when $\alpha$ increases, $B$ increases, goes over a maximum,
and tends to zero for $\alpha\to\pi/2$ ($ \beta$ reaches $\beta_{\rm
max}$).
\par
The Gibbs potential $G({\cal H}_x,{\cal H}_z)$ is now readily found:
\begin{eqnarray}
G&=& {\sqrt{k}\over 4\pi}\,{\cal H}_z
\sqrt{H_0^2 -{\cal H}_z^2}\nonumber\\
&=&{\sqrt{k}\over 4\pi}\,{\cal H}\,\cos\alpha\,
\sqrt{H_0^2 -{\cal H}^2\sin^2\alpha} \,.
\label{GGG}
\end{eqnarray}
In particular, the torque $\tau_y=-dG/d\alpha$ reads:
\begin{equation}
\tau_y ={\sqrt{k}\over 4\pi}\,{\cal H}\,\sin\alpha\,
\frac{H_0^2+{\cal H}^2\cos 2\alpha}
{\sqrt{H_0^2 -{\cal H}^2\sin^2\alpha}}  \,,
\label{trq}
\end{equation}
Note an unusual feature: when $\alpha\rightarrow \pi/2$, the
torque goes to a constant value:
\begin{equation}
\tau (\pi/2) ={{\cal H}\sqrt{k}\over 4\pi}\,\sqrt{H_0^2-{\cal H}^2}.
\label{trq_0}
\end{equation}
\par
For $\alpha>\pi/2$ (or ${\cal H}_z <0$), one has to change sign in Eq.
(\ref{F1}) or to replace ${\cal H}_z$ with $|{\cal H}_z|$. One can
follow the same derivation to see that $|{\cal H}_z|$ appears in Eq.
(\ref{GGG}) for the Gibbs potential. In other words, the potential
$G(\alpha)\propto |\pi/2-\alpha|$ in the small vicinity of
$\alpha=\pi/2$.  A similar situation has been discussed in
Ref.\onlinecite{Pokr}.
\par
The nonanalyticity of $G$ and $\tau$ at $\alpha =\pi/2$ is an artifact of the
infinite slab geometry. To treat better the vicinity of $\alpha =\pi/2$ we
turn to a more realistic sample shape.
\par
\subsection{ Oblate ellipsoid}
The thermodynamic potential which is minimum in equilibrium for this
case is given in Ref. \onlinecite{LL}:
\begin{equation}
\tilde F=F-{\vec{H}\cdot \vec{B}
\over 4\pi}-{1\over 2}\vec{M}\cdot\vec{\cal H}
\label{eq1}
\end{equation}
\par
All macroscopic fields inside the ellipsoid are uniform and related to the
applied field by
\begin{equation}
H_i + n_{ik}(B_k -H_k)={\cal H}_i
\label{eq2} \end{equation}
where $n_{ik}$ is the demagnetization tensor. Let us consider the
sample as an oblate ellipsoid with $z=c$ as the axis of rotation. Then
$n_z=1-2n_x$ and
\begin{eqnarray}
(1-n)H_x+nB_x&=&{\cal H}_x\,,
\label{eq3a}\\
2nH_z+(1-2n)B_z&=&{\cal H}_z\,,
\label{eq3b}
\end{eqnarray}
where the subscript $x$ at $n_x$ is omitted for brevity.
The free energy in this case reads:
\begin{eqnarray}
F&=&{B\over\phi_0}\,\epsilon (\beta)=
{BH_0\over 4\pi }\, \sqrt{\sin^2\beta + k\cos^2\beta}=\nonumber\\
&=&{ H_0\over 4\pi}\,\sqrt{B_x^2+kB_z^2}\,.
\label{eq4}
\end{eqnarray}
The field $H_i=4\pi\,\partial F/\partial B_i$ is now readily found:
\begin{eqnarray}
H_x&=&H_0\,{B_x\over\sqrt{B_x^2+kB_z^2}}\,,
\label{eq74}\\
H_z&=&H_0\,{kB_z\over\sqrt{B_x^2+kB_z^2}}\,.
\label{eq75}
\end{eqnarray}
In principle, Eqs. (\ref{eq3a}), (\ref{eq3b}), (\ref{eq74}), (\ref{eq75})
determine both $\vec{H}$ and $\vec{B}$ in terms of the applied field. This
determination, however, is cumbersome in the general case.
\par
To establish at what field ${\cal H}$ for a given orientation $\alpha$, the
vortices start to penetrate the ellipsoid, we first exclude $B_i$ from the
system (\ref{eq74}), (\ref{eq75}):
\begin{equation}
H_x^2+H_z^2/k=H_0^2\,.\label{**}
\end{equation}
At the Meissner boundary, Eqs. (\ref{eq3a}) and (\ref{eq3b}) yield
\begin{equation}
(1-n)H_x ={\cal H}_x\,,\qquad 2nH_z ={\cal H}_z\,.\label{Meis}
\end{equation}
Therefore,
\begin{equation}
{{\cal H}_x^2\over (1-n)^2} + {{\cal H}_z^2\over
 4n^2k}=H_0^2\,,\label{***}
\end{equation}
or
\begin{equation}
{\cal H}^2\Big[{\sin^2\alpha\over (1-n)^2} + { \cos^2\alpha\over
 4n^2k}\Big]=H_0^2\,.\label{****}
\end{equation}
\par
In the following we restrict ourselves to the case of a flat sample, $n\ll
1$, and to a narrow angular domain near $\alpha=\pi/2$. It is in this domain,
the torque evaluated for an infinite slab ($n_x=0$) has a discontinuous jump.
The simplification here comes about because ${\cal H}_z \ll {\cal H}_x$
along with $B_z\ll B_x$ and $H_z\ll H_x$. Using this and utilizing the
smallness of $n$, we obtain from Eqs. (\ref{eq3a}), (\ref{eq3b}),
(\ref{eq74}), (\ref{eq75}):
\begin{eqnarray}
B_x&\approx &\frac{{\cal H}_x-H_0(1-n)}{n}\,,\quad
B_z\approx \frac{{\cal H}_z}{1-2n}\,, \label{Bxz} \\
 H_x&\approx &H_0\,,\qquad\qquad\qquad\quad
H_z\approx {\cal H}_zn\,\frac{H_0k}{{\cal H}_x-H_0}\,.
\label{Hxz}
\end{eqnarray}
Note that according to Eq. (\ref{****}), for  $\alpha$ close to $\pi/2$, the
field of first penetration is $H_0(1-n)$ and therefore $B_x$ given above is
positive.
\par
It is now a matter of a simple algebra to express the potential
${\tilde F}$ of Eq. (\ref{eq1}) in terms of the applied field:
\begin{equation}
\tilde F\approx {{\cal H}_x(H_0-{\cal H}_x)\over 8\pi n} \, . \label{F-tilde}
\end{equation}
In terms of the angle $\gamma=\alpha-\pi/2$ between the
applied field  and the $ab$ plane, we have
\begin{equation}
\tilde F= {{\cal H}\cos\gamma(H_0-{\cal H}\cos\gamma)\over 8\pi n}
 \,. \label{F-tilde1}
\end{equation}
This yields the torque  for $\gamma\ll 1$:
\begin{equation}
\tau=-{d\tilde F\over d\gamma}= -{{\cal H}\over 8\pi n}
\,  (2{\cal H}-H_0)\, \gamma\,. \label{trq1}
\end{equation}
Thus, as expected, the torque is continuous at $\gamma=0$ ($\alpha =\pi/2$)
where the sample is in the stable equilibrium. The torque is fast increasing
in magnitude when $|\gamma|$ increases.
\par
The crossover between the regime described by Eqs. (\ref{trq}) and
(\ref{trq_0})  takes place around $\gamma_m$ which can be roughly estimated by
equating the torque  given in Eq. (\ref{trq_0}) and that of Eq. (\ref{trq1}):
 \begin{equation}
\gamma_m\sim   2n\sqrt{k}\,.
\label{g_m}
\end{equation}
   The maximum torque then is
\begin{equation}
\tau_m\sim \frac{H_0{\cal H} }{4\pi}\sqrt{k}= \frac{\phi_0k^{1/6}
}{4\pi^3\lambda\lambda_J}\,{\cal H}\,.
\label{tau_m}
\end{equation}
For $\lambda\sim 10^{-5}\,$cm and $\lambda_J\sim 10^{-4}\,$cm, $\tau_m\sim
10^2\,$erg/cm$^3$ in fields on the order $100\,$G. Even for a tiny crystal with
dimensions $(0.1\times 0.1\times 0.01)\,$mm$^3=10^{-7}\,$cm$^3$, we estimate
 $\tau_m$ as $10^{-5}\,$erg. The sensitivity of piezoresistive torque
magnetometers is in the range of  $10^{-7}\,$erg,\cite{Rossel} so that the
torque we have calculated here can, in principle, be measured.
\acknowledgments
This research was supported by grant No. 96-00048 from the
United States - Israel Binational Science Foundation (BSF),
Jerusalem, Israel. The work at Ames was supported by the Office of Basic
Energy Sciences, U.S. Department of Energy. We acknowledge
partial support of the International Institute of Theoretical and
Applied Physics at Iowa State University.
\par

\end{multicols}
\end{document}